\documentclass{sf2a-conf2023}
\usepackage{graphicx}
\usepackage{hyperref}
\usepackage[]{natbib}  
\usepackage{epstopdf}
\usepackage[squaren,Gray]{SIunits}

\def\BibTeX{{\rm B\kern-.05em{\sc i\kern-.025em b}\kern-.08em
    T\kern-.1667em\lower.7ex\hbox{E}\kern-.125emX}}
\bibpunct{(}{)}{;}{a}{}{,}  

\newcommand {\micron}{\unit{}{\micro\meter}}
\newcommand{\HR}{HR\,4796}

\newcommand{\progra}{PROGRA$^2$}

\begin{document}

\TitreGlobal{SF2A 2023}


\title{Insight from laboratory measurements on dust in debris discs}

\runningtitle{The HR\,4796 dust in the lab}





\author{J. Milli}\address{Univ. Grenoble Alpes, CNRS, IPAG, F-38000 Grenoble, France} 
\author{O. Poch$^{1}$}
\author{J.\,B. Renard}\address{LPC2E, Universit\'e d'Orl\'eans, CNRS, Orl\'eans, France} 
\author{J.\,C. Augereau$^{1}$}
\author{P. Beck$^{1}$}
\author{E. Choquet}\address{Aix Marseille Universit\'e, CNRS, CNES,  LAM, Marseille, France} 
\author{J.\,M. Geffrin}\address{Aix Marseille Univ, CNRS, Centrale Marseille, Institut Fresnel, Marseille, France} 
\author{E. Hadamcik}\address{LATMOS, Sorbonne Universit\'e, CNRS, CNES, Paris, France} 
\author{J. Lasue}\address{IRAP, Universit\'e de Toulouse, CNES, CNRS, UPS, Toulouse, Franc} 
\author{F. Menard$^{1}$}
\author{A. Peronne$^{1}$}
\author{C. Baruteau$^{6}$}
\author{R. Tazaki$^{1}$}
\author{V. Tobon\,Valencia$^{1}$}

\setcounter{page}{237}


\maketitle


\begin{abstract}

Extreme adaptive optics instruments have revealed exquisite details on debris discs, allowing to extract the optical properties of the dust particles such as the phase function, the degree of polarisation and the spectral reflectance. These are three powerful diagnostic tools to understand the physical properties of the dust : the size, shape and composition of the dust particles. This can inform us on the population of parent bodies, also called planetesimals, which generate those particles through collisions.
It is however very rare to be able to combine all those three observables for the same system, as this requires different high-contrast imaging techniques to suppress the starlight and reveal the faint scattered light emission from the dust. Due to its brightness, the ring detected around the A-type star \HR{} is a notable exception, with both unpolarised and polarised images covering near-infrared wavelengths. Here, we show how measurements of dust particles in the laboratory can reproduce the observed near-infrared photo-polarimetric properties of the \HR{} disc. Experimental characterisation of dust allows to bypass the current limitations of dust models to reproduce simultaneously the phase function, the degree of polarisation and the spectral reflectance. 

\end{abstract}



\section{Introduction}
  
Debris discs represent an evolutionary stage of a planetary system after a few million years when the primordial gas-rich protoplanetary disc has dissipated, and leaves behind planetesimals that are thought to be leftovers from the planetary formation \cite[see][for a review]{Wyatt2008}. Their mutual collisions produce an optically thin dust disc that is detectable through the infrared excess emission above the stellar photosphere. The initial reservoir of planetesimals depletes with time, hence also the amount of dust. The detection rate of debris discs is indeed $75\%$ in the young 18 Myr $\beta$ Pictoris moving group \citep{Pawellek2021}, but goes down to $\sim20\%$ around Gyr-old stars \cite[e.g.][]{Sibthorpe2018}. 
With an age of $10\pm3$ Myrs, the A-type star \HR{} located at 70.8\,pc \citep{Gaia2020} hosts a particularly well studied debris disc with a large infrared excess of 0.5\% \citep{Moor2006}. Most of the dust is located in a narrow ring at $\sim80$\,au from the central star and a width of about $\sim10$ au \citep{Kennedy2018_HR4796}. Because of its relatively high surface brightness compared to other debris discs, this system has been extensively studied in many wavelengths, as visible in Fig. \ref{fig_images_disc}.

Despite this extensive spectral coverage, the composition, shape and size of the particles remain a mystery since the high-contrast imagers VLT/SPHERE and Gemini/GPI have revealed the scattering phase function (hereafter SPF) and degree of polarisation (hereafter DoLP) of the particules \citep{Perrin2015,Milli2017}. We will focus here on the near-infrared scattering properties of the disc to show how laboratory measurements of dust analogues can be used to shed light on the nature of the particles in this system.
 
\begin{figure}[ht!]
 \centering
 \includegraphics[width=0.9\textwidth,clip]{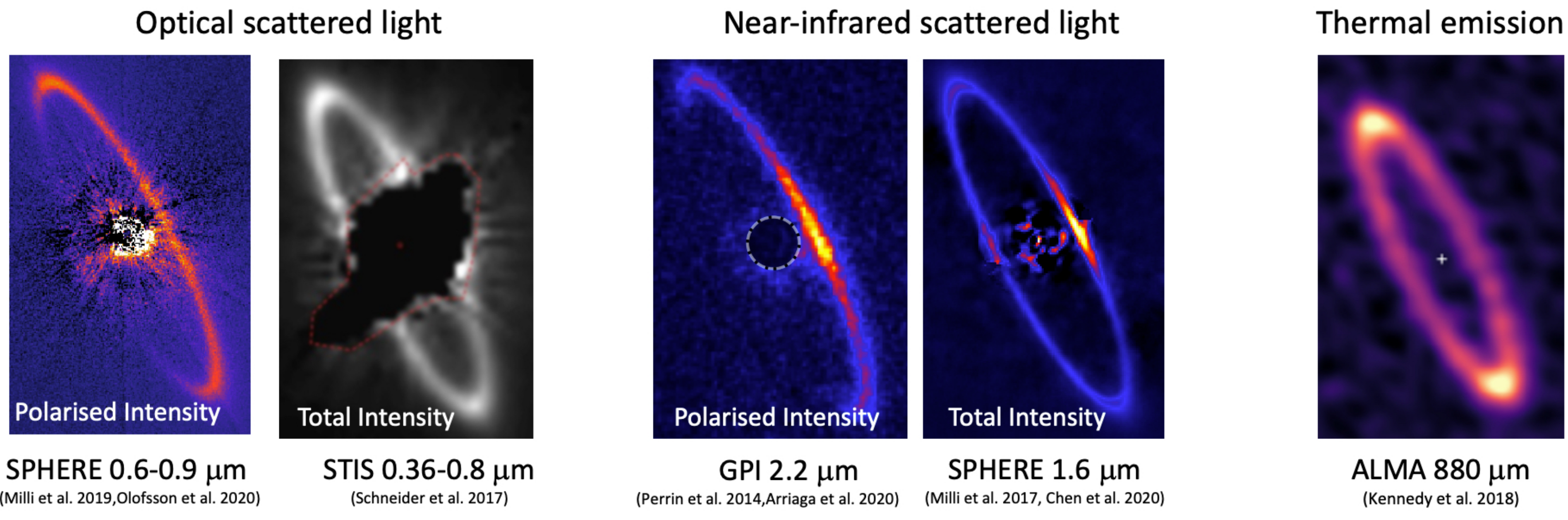}      
  \caption{The \HR{} disc at various wavelengths. The semi-major axis of the disc is $\sim1$". North is up, East to the left.}
  \label{fig_images_disc}
\end{figure}

\section{The scattered light near-infrared properties of the dust}

Thanks to the polarimetric capabilities of the high-contrast instruments SPHERE \citep{Beuzit2018} and GPI \citep{Macintosh2014}, the surface brightness of the ring can be measured at almost all azimuth angles, even along the minor axis. Each position along the ring corresponds to a unique angle between the illumination direction and the line of sight, called the scattering angle. The ring is inclined by $\sim76.5^\circ$ from pole-on, meaning that the minor axes correspond to scattering angles of $13.5^\circ$  and $166.5^\circ$ respectively for the bright North-West forward-scattering side, and the faint South-East backward scattering side. One can therefore extract the SPF and DoLP of the dust particles in this range of scattering angles, as shown in Fig \ref{fig_images_disc}.

\begin{figure}[ht!]
 \centering
 \includegraphics[width=0.9\textwidth,clip]{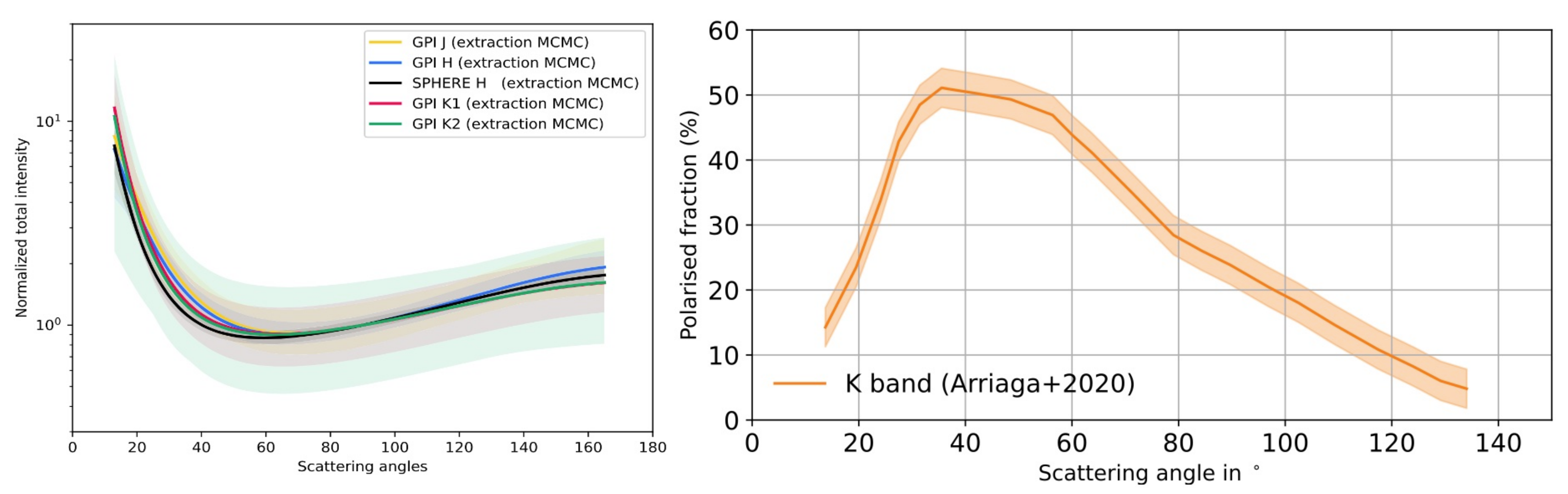}      
  \caption{\textbf{Left:} Total intensity scattering phase function of the disc, as extracted from GPI and SPHERE using a parametric modelling (double-component Heyney Greenstein model) in the J ($1.2\micron$), H ($1.6\micron$) and K ($2.2\micron$) band \citep{Chen2020}. \textbf{Right:} Degree of linear polarisation as extracted from GPI in the K band \citep{Arriaga2020}}
  \label{fig_images_disc}
\end{figure}

Modelling of the dust particles assuming compact spheres (Mie theory) made of basic cometary-like materials used traditionally for debris disc,  e.g. silicates, amorphous carbon, water ice and porosity \citep[see for instance][]{Augereau1999}, could not reproduce the total intensity scattering phase function of the dust particles \citep{Milli2017}. If one relaxes the assumption on the composition of the particles, \citet{Chen2020} showed that the total intensity SPF can be fit using the Distribution of Hollow Spheres \cite[DHS,][]{Min2005} but required a relatively high volume fraction of metallic iron of $37^{+15}_{9}\%$, a highly absorbing mineral, in addition to amorphous carbon and silicates. 
However, when adding the DoLP as a constraint, no model using the Mie nor the DHS theory could simultaneously reproduce the SPF and the DoLP \citep{Arriaga2020}. This points out that one of our initial assumption is wrong, either particles in the \HR{} disc cannot be modelled by compact spheres, as suggested already in \citet{Milli2014}, or a simple differential power-law size distribution is too simplistic. We decided to use laboratory experiments rather than numerical experiments to find good analogues to the \HR{} dust. The choice of the sample was driven by the fact that there are few micron-size particles with a peak in DoLP at scattering angles as small as $40^\circ$, while larger mm-size pebbles can do so \cite[e.g.][]{Munoz2021} but are incompatible with the spectral energy distribution of the disc that requires a minimum particle size of $\sim2\micron$. We found an interesting sample in the \progra{} database\footnote{\url{https://www.icare.univ-lille.fr/progra2-en/?noredirect=en_US}}, pyrrhotite, a non-stoichiometric form of iron sulphide Fe$_{1-x}$S ($0<x\leq 0.125$).

\section{Laboratory measurements of dust particles with \progra}

\subsection{The \progra{} instrument and pyrrhotite sample}

\begin{figure}[ht!]
 \centering
 \includegraphics[width=0.8\textwidth,clip]{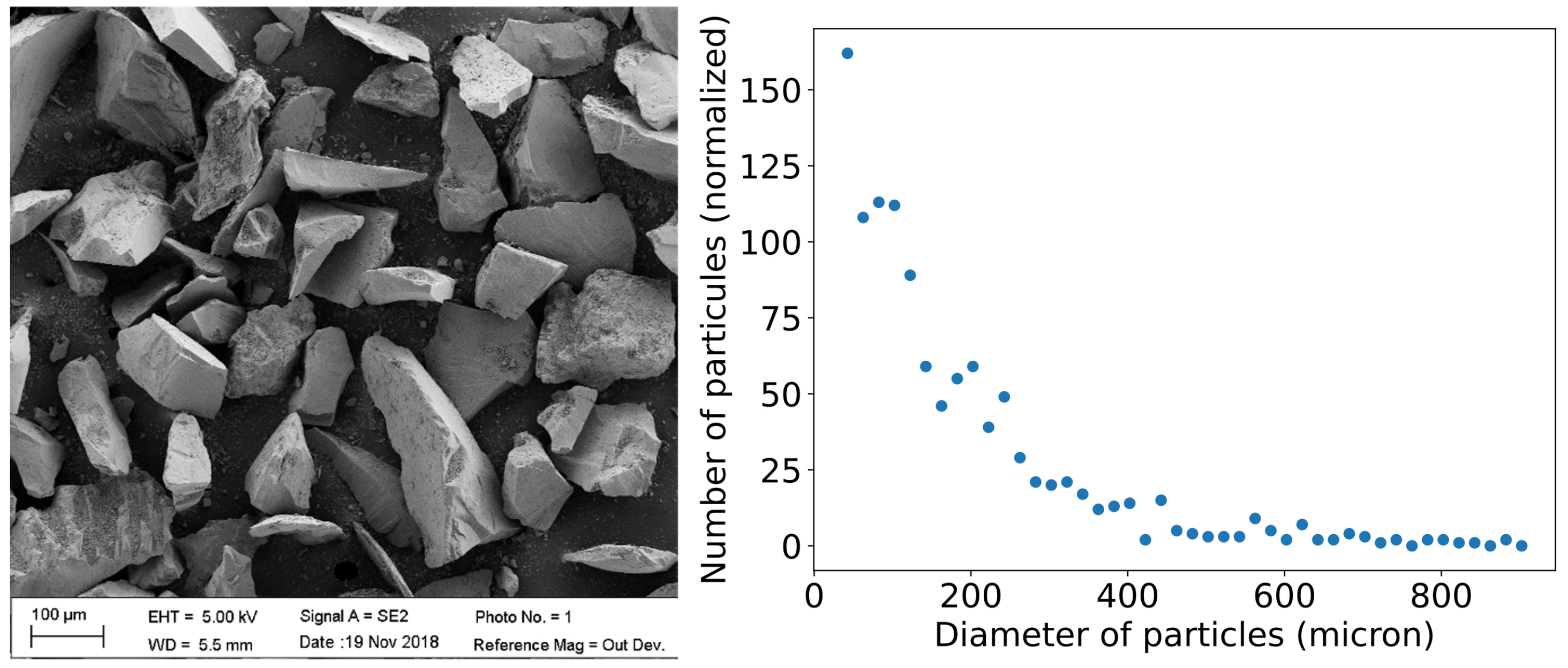}      
  \caption{\textbf{Left:} Scanning Electron Microscope view of the sample . \textbf{Right:} Granulometry of the sample as measured using the camera onboard \progra{} in the R band. The resolution of the camera is limited to $40\micron$.}
  \label{fig_pyrrhotite}
\end{figure}

The \progra{} instrument is dedicated to the study of light scattered by solid particles deposited on a surface or lifted in clouds \citep{Worms2000,Hadamcik2002}. We used here a sample of pyrrhotite (Fig. \ref{fig_pyrrhotite}) that was lifted in clouds in microgravity conditions during parabolic flights that took place between 2018 and 2021. The sample is contained in a vial where it receives the light from a laser. The light scattered by the particles is split using a polarised beam splitter in two channels where it is measured on two detectors. 

\subsection{Measurements and conclusions}

\begin{figure}[ht!]
 \centering
 \includegraphics[width=0.8\textwidth,clip]{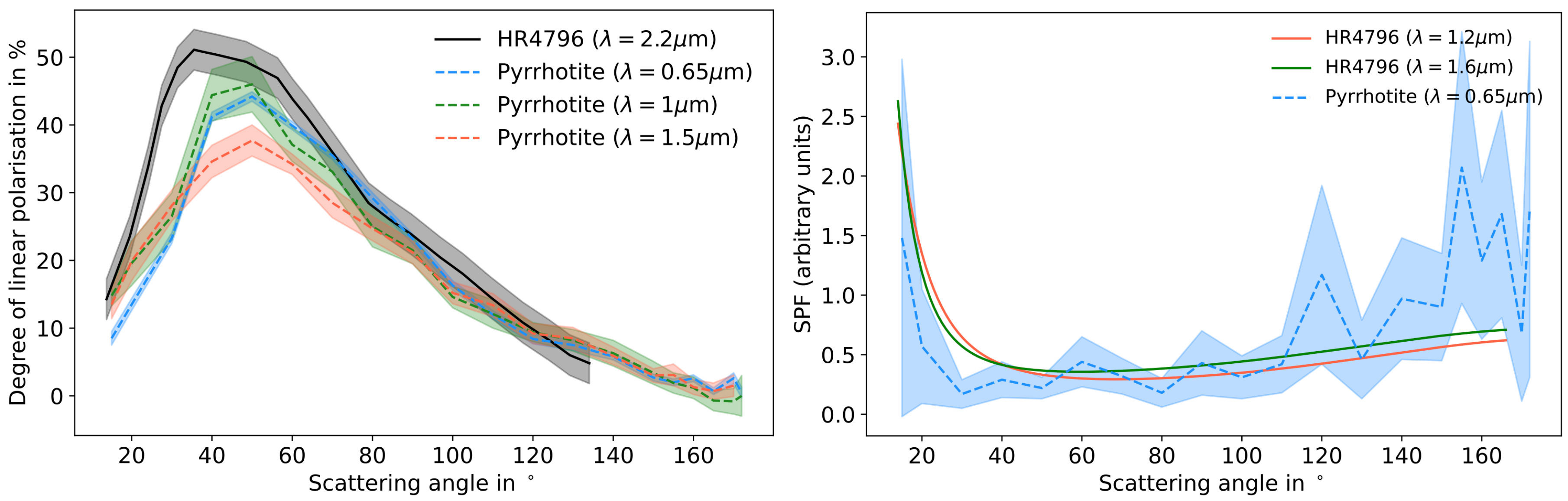}      
\caption{\textbf{Left:} DoLP of the measured pyrrhotite sample compared to the DoLP of HR4796 in the K band \citep{Arriaga2020}. \textbf{Right:} Optical SPF of the pyrrhotite sample compared to the NIR SPF of HR4796  \citep{Chen2020}}
  \label{fig_progra2_measurements}
\end{figure}

\progra{} allows measurements of the total intensity SPF at optical wavelengths, and the DoLP at both optical and NIR wavelengths. The comparison to the \HR{} particles are shown in Fig. \ref{fig_progra2_measurements}. We notice that the peak in DoLP around a scattering angle of $40^\circ$ is well reproduced by the sample. There are very few samples with particles with sizes in the 1-200$\micron$ range that can reproduce a peak at so small scattering angles in the large \progra{} database: only silicon carbides (CSi) could also produce this behaviour, with an overall match not as good as pyrrhotite.  The maximum DoLP as measured in the optical and near-infrared (NIR) is in the range 35-45\% for the 1-200$\micron$ pyrrhotite, slightly smaller than the \HR{} particles at $2.2\micron$, but the maximum DoLP can increase with wavelength for large (35-150$\micron$) levitating grains \citep{Renard2014} and can also largely dependent on the minimum particle size or size distribution (Milli et al. in prep.). Regarding the total intensity SPF (Fig. \ref{fig_progra2_measurements} right), the backward scattering behaviour seen on \HR{} in the NIR is well reproduced by the pyrrhotite sample measured in the optical. Unfortunately, \progra{} does not have the capability to measure the SPF in the NIR for a rigorous comparison at the same wavelength.

Overall, the pyrrhotite sample is an interesting candidate material to reproduce simultaneously the total intensity SPF and the DoLP of the particles in \HR. In this study, no attempt to adjust the sample to match exactly the \HR{} properties has been made, in particular the particle sizes, but we present here for the first time, archival \progra{} measurements done on a sample of pyrrhotite with sizes dominated by particles smaller than $100\micron$. The presence of a highly absorbing material such as iron sulphide confirms the finding of \citet{Chen2020} using the total intensity SPF alone and bring guidance in the attempt to reconcile the SPF and DoLP with more irregular particles than the compact spheres tested in \citet{Arriaga2020}. The presence of iron sulphides would not be surprising in circumstellar dust.  
Stratospheric interstellar dust particles, Antarctic Micro-Meteorites, comet Wild 2 samples, all contain sulphides in the form of troilite FeS, among other \citep{Dobrica2009}. Iron is also present in the comet 67P/Churyumov-Gerasimenko, with a mass fraction of 7.5\% \citep{Bardyn2017}. It is probably mostly present in the form of iron sulphides and Fe-Ni alloys, which are opaque minerals responsible for the dark reflectance from optical to NIR wavelengths of cometary and primitive asteroids surfaces \citep{Quirico2016}, they could therefore also be present on the surface of the \HR{} particles, explaining at the same time the red reflectance spectrum of the disc \citep{Milli2017}.

\begin{acknowledgements}

We dedicate this paper to the memory of Anny-Chantal Levasseur-Regourd (ACLR), who encouraged and inspired our efforts to compare remote observations of exoplanetary dust to laboratory measurements. This work was carried out as part of the EPOPEE project (Etude des POussi\`eres Plan\'etaires Et Exoplan\'etaires) project. This research group is funded by the Programme National de Plan\'etologie (PNP) of CNRS-INSU in France, from which we acknowledge financial support. ACLR was one of the most enthusiastic members, always willing to share her wide expertise with the younger generation, and she has vastly contributed to this scientific collaboration between the disc and solar system community. She sadly passed away only a few weeks after attending our June 2022 EPOPEE working group meeting in Grenoble. 

\end{acknowledgements}

\bibliographystyle{aa}  
\bibliography{biblio} 

\end{document}